\pdfoutput=1

\documentclass[twocolumn,aps,prl,superscriptaddress]{revtex4}

\usepackage{times}
\usepackage{graphicx}
\usepackage{epsfig}
\usepackage{amsfonts}
\usepackage{amsmath}
\usepackage{amssymb}
\usepackage{color}
\usepackage{multirow}
\usepackage[colorlinks=true,citecolor=blue,urlcolor=blue]{hyperref}

\begin{document}


\title{Exclusivity principle and the quantum bound of the Bell inequality}


\author{Ad\'an~Cabello}
 \email{adan@us.es}
 \affiliation{Departamento de F\'{\i}sica Aplicada II, Universidad de
 Sevilla, E-41012 Sevilla, Spain}


\date{\today}


\begin{abstract}
 We show that, for general probabilistic theories admitting sharp measurements, the exclusivity principle together with two assumptions exactly singles out the Tsirelson bound of the Clauser-Horne-Shimony-Holt Bell inequality.
\end{abstract}



\maketitle


{\em Introduction.---}Quantum theory (QT) is arguably the most accurate scientific theory of all times. Nevertheless, despite its mathematical simplicity, its fundamental principles are still unknown. In the quest for these principles, a key question is why QT is exactly as nonlocal \cite{PR94} and contextual \cite{Cabello11} as it is.

There are two different approaches to answer this question. On one hand, the ``black box'' approach, which focuses on correlations among the outcomes of measurements in multipartite scenarios. This approach makes no assumptions on the internal working of the measurement devices (which are treated as black boxes) and pays no attention to the postmeasurement state of the system (since measurements are treated as if they were demolition measurements). Within this approach, it has been proven that the principles of information causality \cite{PPKSWZ09} and macroscopic locality \cite{NW09}, complemented by some assumptions, single out the maximum quantum violation of the Clauser-Horne-Shimony-Holt (CHSH) Bell inequality \cite{Bell64,CHSH69} (i.e., the Tsirelson bound \cite{Tsirelson80}). However, it has also been proven that these principles cannot exclude extremal nonlocal boxes prohibited in QT in the three-party, two-outcome, two-setting or $(3,2,2)$ scenario \cite{GWAN11,YCATS12}. Moreover, by its very definition, the black box approach cannot explain quantum contextual correlations \cite{Cabello11}. In addition, there is increasing evidence that the black box approach, with independence of the principle invoked, cannot explain even quantum nonlocal correlations \cite{NGHA14}.

On the other hand, the ``sharp measurements,'' ``graph-theoretic,'' or ``contextuality'' approach \cite{CSW10, CSW14, RDLTC14} focus on correlations among the outcomes of sharp measurements. For general probabilistic theories (GPTs),
sharp measurements are nondemolition measurements that are repeatable and cause minimal disturbance \cite{Kleinmann14, CY14}. In QT, sharp measurements are projective measurements \cite{vonNeumann32}. In QT, generalized measurements are represented by positive operator-valued measures. However, in QT, any generalized measurement can always be implemented as a sharp measurement on the system and the environment \cite{GN43}.

Within the sharp measurements approach, an ``event'' is defined as the postmeasurement state of the system after a sharp measurement. Two events are equivalent when they correspond to indistinguishable states. Two events are exclusive when there is a sharp measurement that perfectly distinguishes between them.

The exclusivity (E) principle \cite{Cabello13,Yan13,CDLP13,FSABCLA13,ATC14,SFABCLA14} states that any set of $m$ pairwise exclusive events is $m$-wise exclusive. According to Kolmogorov's axioms of probability, the sum of the probabilities of $m$ $m$-wise exclusive events cannot exceed~1. Therefore, assuming the E principle, the sum of the probabilities of any set of pairwise exclusive events cannot exceed~1. Notice, however, that the E principle is not implied by Kolmogorov's axioms. As noticed by Specker, there are theories satisfying Kolmogorov's axioms which do not satisfy the E principle \cite{Specker60} (see also Ref.~\cite{LSW11}). Later Specker conjectured that the E principle could be the fundamental principle of QT \cite{Specker09}. Recent works have shown that the E principle is implied by physical principles of a very different nature \cite{CY14, BMU14, Henson14}.

So far, the E principle, plus some assumptions, has been proven to single out the maximum quantum violation of the simplest noncontextuality inequality \cite{Cabello13}, to exclude all extremal nonlocal boxes in the $(3,2,2)$ scenario \cite{FSABCLA13,SFABCLA14}, and to exclude any set of correlations strictly larger than the set of correlations achievable with quantum systems for any self-complementary graph of exclusivity \cite{ATC14}. There is also strong evidence that the E principle singles out the maximum quantum violation of all basic noncontextuality inequalities \cite{CDLP13}. However, so far, the E principle has failed to explain the maximum quantum violation of the CHSH Bell inequality (although it has provided upper bounds \cite{Cabello13,FSABCLA13,SFABCLA14}). Here we will show that the E principle, plus some assumptions, exactly singles out the Tsirelson bound of the CHSH Bell inequality.


{\em The Tsirelson bound.---}The discovery that QT violates the CHSH Bell inequality, a condition that any local realistic (LR) theory must satisfy, is one of the most celebrated results of science \cite{Bell64}. The CHSH Bell inequality \cite{CHSH69} is defined in a scenario in which there are two distant observers, Alice and Bob, and in each run each of them measures an observable randomly chosen between two. Alice chooses between $A_0$ and $A_1$, and Bob between $B_0$ and $B_1$. Each measurement has two possible results: $+1$ and $-1$. Alice's (Bob's) choice is spacelike separated from Bob's (Alice's) measurement result. This implies that one observer's result cannot be influenced by the other observer's choice, assuming that influences do not propagate faster than the speed of light in vacuum.

The CHSH Bell inequality states that, for any LR theory,
\begin{equation}
 S \stackrel{\mbox{\tiny{LR}}}{\leq} 3,
\end{equation}
where
\begin{eqnarray}
 S=&&P(A_0+,B_0+)+P(A_0-,B_0-)+P(A_0+,B_1+) \nonumber \\
 &&+P(A_0-,B_1-)+P(A_1+,B_0+)+P(A_1-,B_0-) \nonumber \\
 &&+P(A_1+,B_1-)+P(A_1-,B_1+),
 \label{S}
\end{eqnarray}
where, e.g., $P(A_1+,B_1-)$ is the joint probability of Alice obtaining $+1$ when measuring $A_1$ and Bob obtaining $-1$ when measuring $B_1$. However, according to QT,
\begin{equation}
 S \stackrel{\mbox{\tiny{QT}}}{\leq} 2 + \sqrt{2} \approx 3.414.
\end{equation}
This upper bound is the Tsirelson bound \cite{Tsirelson80}. Assuming that QT is correct, this means that the universe is ``nonlocal'' (i.e., it cannot be explained with LR theories), but only up to a certain limit. An intriguing question is why \cite{PR94}.


{\em The E principle singles out the Tsirelson bound.---}To make the proof simple, let us begin by assuming that the maximum of $S$ is attained when each of the eight joint probabilities in (\ref{S}) takes the same value $p$. We will later remove this assumption. Given the eight joint probabilities in (\ref{S}), the conditions of normalization of the probabilities and the conditions of comeasurability in the CHSH Bell scenario determine the values of the eight complementary probabilities $P(A_0+,B_0-),\ldots, P(A_1-,B_1-)$. In particular, if each of the eight joint probabilities in (\ref{S}) equals $p$, then each of the eight complementary probabilities have to be $\frac{1}{2}-p$. To determine the maximum value of $p$ we proceed in four steps.

(I) Let us consider two CHSH Bell experiments, one performed in one city, e.g., Vienna, on pairs of particles prepared in a certain state and another experiment in another city, e.g., Stockholm, on different pairs of particles also prepared in the same state. In Vienna, Alice randomly measures $A_0$ or $A_1$, and Bob randomly measures $B_0$ or $B_1$. In Stockholm, Alice$'$ randomly measures $A_0'$ or $A_1'$ (which are the same ones Alice is measuring in Vienna), and Bob$'$ randomly measures $B_0'$ or $B_1'$ (which are the same ones Bob is measuring in Vienna).
We make the assumption that the Vienna and Stockholm experiments are independent in the sense that the joint probability of an event involving Vienna's and Stockholm's events is the product of the probabilities of the respective events. That is,
\begin{equation}
 P(A_ia,B_jb,A'_ka',B'_lb')=P(A_ia,B_jb) P(A'_ka',B'_lb').
 \label{factorization}
\end{equation}
Therefore, e.g., $P(A_0+,B_0+,A'_0+,B'_1+)=p^2$ and $P(A_0+,B_0-,A'_0+,B'_1-)=\left(\frac{1}{2}-p\right)^2$.


\begin{figure}[tb]
\centering
\vspace{-3cm}
\hspace{-8mm}
\includegraphics[scale=0.44]{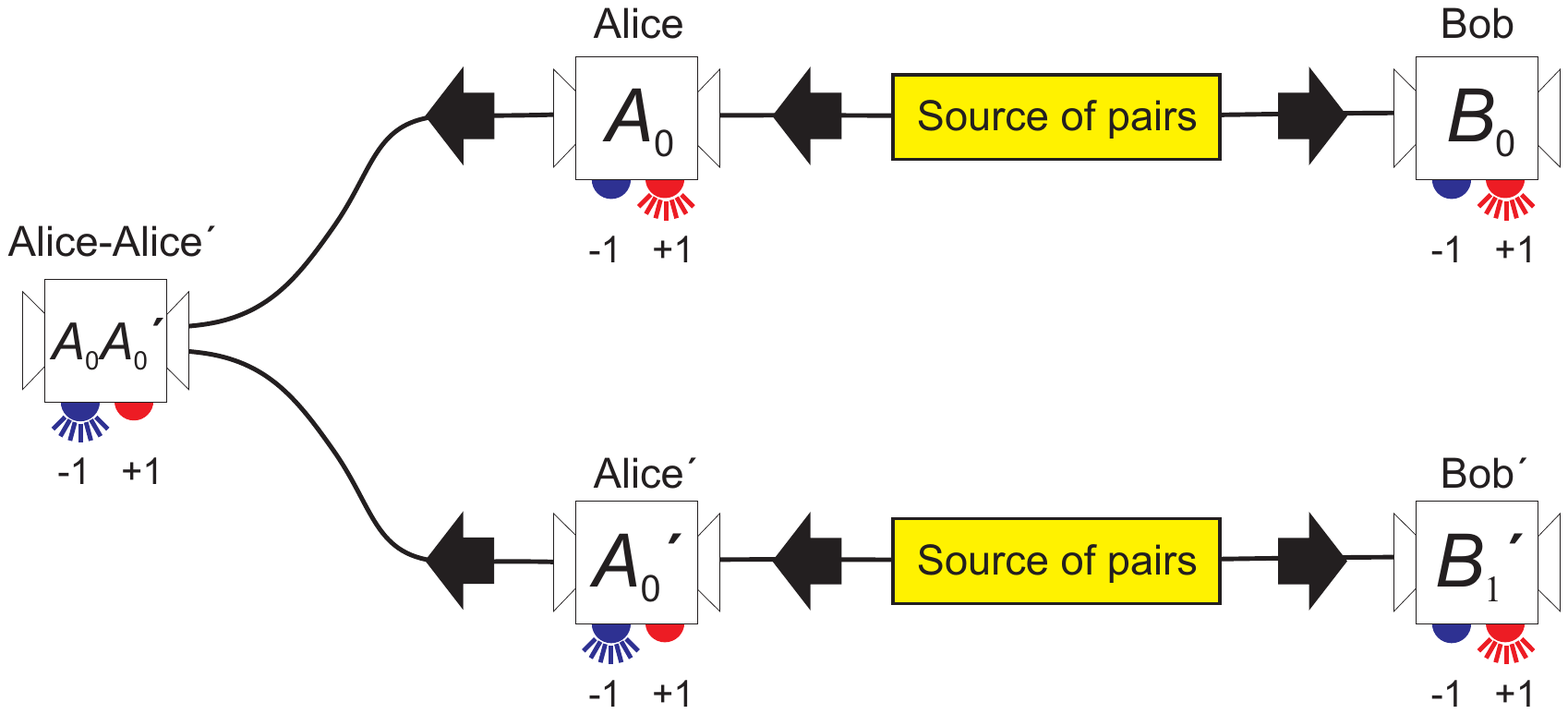}
\vspace{-5.6cm}
\caption{\label{Fig1} (Color online) A key point of the proof is the observation that extra measurements may be carried out after sharp measurements used for a CHSH Bell experiment. Specifically, after Alice has measured $A_i$ and Alice$'$ has measured $A'_i$, Alice and Alice$'$ can also measure $A_iA'_i$.}
\end{figure}


(II) Now recall that we are assuming that the measurements are sharp. Therefore, in principle, extra measurements may be carried out after those needed for the CHSH Bell experiment. In particular, let us consider the measurements $A_iA'_i$, with $i=0,1$, defined as follows: $A_iA'_i$ is comeasurable with $A_i$ and $A'_i$ and gives the result $A_iA'_i-$ if $A_i+$ and $A'_i-$ or if $A_i-$ and $A'_i+$, and gives $A_iA'_i+$ if $A_i+$ and $A'_i+$ or if $A_i-$ and $A'_i-$ (see Fig.~\ref{Fig1}). Notice that the existence of $A_iA'_i$ with these properties is guaranteed if $A_i$ and $A'_i$ are sharp and comeasurable, as we are assuming here. Notice also that the events $(A_0+,B_0+,A'_0+,B'_1+)$ and $(A_0+,B_0+,A'_0+,B'_1+,A_0A'_0+)$ are equivalent. Therefore, $P(A_0+,B_0+,A'_0+,B'_1+)=P(A_0+,B_0+,A'_0+,B'_1+,A_0A'_0+)$.

(III) Now let us assume that there is a joint probability distribution for $A_0A'_0$ and $A_1A'_1$ (which, remember, are defined from $A_0$ and $A_1$) when $A_0$ and $A_1$ give the maximum of $S$. Then,
\begin{equation}
 \sum_{a,b\in\{+,-\}} P(A_0A'_0a, A_1A'_1b)=1.
 \label{assumption}
\end{equation}


\begin{table*}[hbt]
\caption{Four sets of nine pairwise exclusive events and their probabilities.\label{Table1}}
\begin{tabular}{cc|cc}
\hline
\hline
Event & Probability & Event & Probability \\ [0.5ex]
\hline
$e_1=(A_0+,B_0+,A'_0+,B'_1+,A_0A'_0+)$ & $p^2$ & $f_1=(A_0+,B_0+,A'_0+,B'_0+,A_0A'_0+)$ & $p^2$ \\
$e_2=(A_0-,B_0-,A'_0-,B'_1-,A_0A'_0+)$ & $p^2$ & $f_2=(A_0-,B_0-,A'_0-,B'_0-,A_0A'_0+)$ & $p^2$ \\
$e_3=(A_0+,B_0-,A'_0+,B'_1-,A_0A'_0+)$ & $\left(\frac{1}{2}-p\right)^2$ & $f_3=(A_0+,B_0-,A'_0+,B'_0-,A_0A'_0+)$ & $\left(\frac{1}{2}-p\right)^2$ \\
$e_4=(A_0-,B_0+,A'_0-,B'_1+,A_0A'_0+)$ & $\left(\frac{1}{2}-p\right)^2$ & $f_4=(A_0-,B_0+,A'_0-,B'_0+,A_0A'_0+)$ & $\left(\frac{1}{2}-p\right)^2$ \\
$e_5=(A_1+,B_0+,A'_1+,B'_1-,A_1 A'_1+)$ & $p^2$ & $f_5=(A_1+,B_0+,A'_1-,B'_0-,A_1 A'_1-)$ & $p^2$ \\
$e_6=(A_1-,B_0-,A'_1-,B'_1+,A_1 A'_1+)$ & $p^2$ & $f_6=(A_1-,B_0-,A'_1+,B'_0+,A_1 A'_1-)$ & $p^2$ \\
$e_7=(A_1+,B_0-,A'_1+,B'_1+,A_1 A'_1+)$ & $\left(\frac{1}{2}-p\right)^2$ & $f_7=(A_1+,B_0-,A'_1-,B'_0+,A_1 A'_1-)$ & $\left(\frac{1}{2}-p\right)^2$ \\
$e_8=(A_1-,B_0+,A'_1-,B'_1-,A_1 A'_1+)$ & $\left(\frac{1}{2}-p\right)^2$ & $f_8=(A_1-,B_0+,A'_1+,B'_0-,A_1 A'_1-)$ & $\left(\frac{1}{2}-p\right)^2$ \\
$e_9=(A_0A'_0-,A_1 A'_1-)$ & $P(e_9)$ & $f_9=(A_0A'_0-,A_1 A'_1+)$ & $P(f_9)$ \\
\hline
Event & Probability & Event & Probability \\ [0.5ex]
\hline
$g_1=(A_0+,B_0+,A'_0-,B'_0-,A_0A'_0-)$ & $p^2$ & $h_1=(A_0+,B_0+,A'_0-,B'_1-,A_0A'_0-)$ & $p^2$ \\
$g_2=(A_0-,B_0-,A'_0+,B'_0+,A_0A'_0-)$ & $p^2$ & $h_2=(A_0-,B_0-,A'_0+,B'_1+,A_0A'_0-)$ & $p^2$ \\
$g_3=(A_0+,B_0-,A'_0-,B'_0+,A_0A'_0-)$ & $\left(\frac{1}{2}-p\right)^2$ & $h_3=(A_0+,B_0-,A'_0-,B'_1+,A_0A'_0-)$ & $\left(\frac{1}{2}-p\right)^2$ \\
$g_4=(A_0-,B_0+,A'_0+,B'_0-,A_0A'_0-)$ & $\left(\frac{1}{2}-p\right)^2$ & $h_4=(A_0-,B_0+,A'_0+,B'_1-,A_0A'_0-)$ & $\left(\frac{1}{2}-p\right)^2$ \\
$g_5=(A_1+,B_0+,A'_1+,B'_0+,A_1 A'_1+)$ & $p^2$ & $h_5=(A_1+,B_0+,A'_1-,B'_1+,A_1 A'_1-)$ & $p^2$ \\
$g_6=(A_1-,B_0-,A'_1-,B'_0-,A_1 A'_1+)$ & $p^2$ & $h_6=(A_1-,B_0-,A'_1+,B'_1-,A_1 A'_1-)$ & $p^2$ \\
$g_7=(A_1+,B_0-,A'_1+,B'_0-,A_1 A'_1+)$ & $\left(\frac{1}{2}-p\right)^2$ & $h_7=(A_1+,B_0-,A'_1-,B'_1-,A_1 A'_1-)$ & $\left(\frac{1}{2}-p\right)^2$ \\
$g_8=(A_1-,B_0+,A'_1-,B'_0+,A_1 A'_1+)$ & $\left(\frac{1}{2}-p\right)^2$ & $h_8=(A_1-,B_0+,A'_1+,B'_1+,A_1 A'_1-)$ & $\left(\frac{1}{2}-p\right)^2$ \\
$g_9=(A_0A'_0+,A_1 A'_1-)$ & $P(g_9)$ & $h_9=(A_0A'_0+,A_1 A'_1+)$ & $P(h_9)$ \\
\hline
\hline
\end{tabular}
\end{table*}


(IV) Now let us consider the four sets nine pairwise exclusive events given in Table~\ref{Table1}. Since each set $\{e_i\}$, $\{f_i\}$, $\{g_i\}$, and $\{h_i\}$, with $i=1,\ldots, 9$, is a set of pairwise exclusive events, the E principle enforces
\begin{equation}
\sum_{i=1}^{9} P(e_i) \stackrel{\mbox{\tiny{E}}}{\leq} 1,\;\sum_{i=1}^{9} P(f_i) \stackrel{\mbox{\tiny{E}}}{\leq} 1,\;
\sum_{i=1}^{9} P(g_i) \stackrel{\mbox{\tiny{E}}}{\leq} 1,\;\sum_{i=1}^{9} P(h_i) \stackrel{\mbox{\tiny{E}}}{\leq} 1.
\label{Erestrictions}
\end{equation}
Summing these four inequalities we obtain
\begin{equation}
\sum_{i=1}^{9} P(e_i)+P(f_i)+P(g_i)+P(h_i)\stackrel{\mbox{\tiny{E}}}{\leq} 4.
\end{equation}
Taking into account the probabilities given in Table~\ref{Table1} and Eq.~(\ref{assumption}), we obtain
\begin{equation}
p \stackrel{\mbox{\tiny{E}}}{\leq} \frac{2+\sqrt{2}}{8}.
\end{equation}
Therefore,
\begin{equation}
S \stackrel{\mbox{\tiny{E}}}{\leq} 2 + \sqrt{2}.
\label{fin}
\end{equation}

If the eight joint probabilities in (\ref{S}) take arbitrary values, notice that there are only two sets of nine mutually exclusive events containing $(A_0A'_0-,A_1A'_1-)$ like the one in Table~\ref{Table1}, i.e., with eight events $(A_ia,B_jb,A'_ka',B'_lb')$ such that in four of them $(A_ia,B_jb)$ and $(A'_ka',B'_lb')$ are in (\ref{S}) and in the other four are in the complementary set. Consider the two sets corresponding to each of $(A_0A'_0\pm,A_1A'_1\pm)$, $(A_0A'_0\pm,A_1A'_1\mp)$, $(A_0A'_1\pm,A_1A'_0\pm)$, and $(A_0A'_1\pm,A_1A'_0\mp)$. For each set, the E principle enforces a restriction like the ones in (\ref{Erestrictions}). Summing all of them and considering (\ref{assumption}) and its symmetric version, namely, $\sum_{a,b\in\{+,-\}} P(A_0A'_1a, A_0A'_1b)=1$, we obtain
\begin{equation}
S^2+(4-S)^2 + 4 \stackrel{\mbox{\tiny{E}}}{\leq} 16
\end{equation}
and we obtain, again, inequality (\ref{fin}). This finishes the proof.


{\em Discussion.---}The proof makes no reference to QT at all. It is valid for all GPTs admitting sharp measurements satisfying the E principle and under the two assumptions made in steps (I) and (III). The assumption in step (I) has been made in other works (e.g., in Refs. \cite{CDLP13, FSABCLA13}).

The assumption in step (III) holds for GPTs different than QT. Of course, it also holds in QT for sharp measurements
and for sharp extensions of generalized measurements. It may be the case that this assumption is not needed when an infinite number of copies of the CHSH Bell experiment are considered. However, proving this conjecture is related to an open problem in graph theory \cite{Cabello13, FSABCLA13}. Nevertheless, the proof presented here has the virtue that, unlike other proofs \cite{PPKSWZ09,NW09}, does not require an infinite number of copies of the CHSH Bell experiment and, consequently, an infinite universe \cite{MM14}.

On the other hand, the fact that the maximum violation of the CHSH Bell inequality is singled out by the E principle (under the assumptions mentioned) implies (by Result~2 in Ref.~\cite{ATC14}) that the maximum quantum violation of another noncontextuality inequality is also singled out by the E principle (under the assumptions mentioned) \cite{NBDASBC13}.


{\em Conclusions.---}Until now, the E principle failed to explain the maximum violation of the CHSH Bell inequality. Here we have presented such an explanation (under the assumptions mentioned). The problem of whether or not the E principle (complemented by some assumptions) can explain all the quantum limits of the correlations between the outcomes of comeasurable sharp measurements is still open. However, taking into account the earlier successful predictions of the E principle \cite{Cabello13,Yan13,CDLP13,FSABCLA13,ATC14,SFABCLA14} and the evidences of failure of other approaches \cite{GWAN11,YCATS12,NGHA14}, the result presented here promotes the E principle as the best option for understanding the limits of quantum contextual and nonlocal correlations. This apparent fundamental role of the E principle in QT is also supported by some recent results aiming to understand QT from fundamental physical principles \cite{BMU14,CY14,Henson14}.




{\em Acknowledgments.---}The author thanks L.\ Hardy, M.\ Navascu\'es, and M.\ Paw{\l}owski for useful conversations, and M.\ Kleinmann and J.-\AA.\ Larsson for suggestions on the manuscript. This work was supported by the project FIS2011-29400 (MINECO, Spain) with FEDER funds, the FQXi large grant project ``The Nature of Information in Sequential Quantum Measurements,'' and the program ``Science Without Borders'' (CAPES and CNPq, Brazil).



\end{document}